\documentstyle[prl,aps,floats,twocolumn,epsfig]{revtex}

\newcommand{\beq}{\begin{equation}}
\newcommand{\eeq}{\end{equation}}
\newcommand{\bea}{\begin{eqnarray}}
\newcommand{\eea}{\end{eqnarray}}

\def\laq{~\raise 0.4ex\hbox{$<$}\kern -0.8em\lower 0.62
ex\hbox{$\sim$}~}
\def\gaq{~\raise 0.4ex\hbox{$>$}\kern -0.7em\lower 0.62
ex\hbox{$\sim$}~}

\def \ra {\rightarrow}

\def \Da {\Delta}

\def \ga {\gamma}

\def \da {\delta}

\def \Om {\Omega}

\def \ti {\tilde}

\def \mt {\tilde m}

\begin{document}

\par
\begingroup
\twocolumn[%

\begin{flushright}
BA-TH/01-421\\
gr-qc/0107103\\
\end{flushright}
\bigskip

{\large\bf\centering\ignorespaces
Sensitivity of spherical gravitational-wave detectors to a stochastic
background \\of non-relativistic scalar radiation 

\vskip2.5pt}
{\dimen0=-\prevdepth \advance\dimen0 by23pt
\nointerlineskip \rm\centering
\vrule height\dimen0 width0pt\relax\ignorespaces
E. Coccia${}^{(1,2)}$, M. Gasperini${}^{(3,4)}$ and C. 
Ungarelli${}^{(5)}$ 
\par}
{\small\it\centering\ignorespaces
${}^{(1)}$
Dipartimento di Fisica, Universit\`a di Roma 2 
``Tor Vergata",  Via della Ricerca Scientifica 1, 00133 Roma, Italy \\
${}^{(2)}$
Istituto Nazionale di Fisica Nucleare, Sezione di Roma 2,
Roma, Italy \\
${}^{(3)}$
Dipartimento di Fisica, Universit\`a di Bari, 
Via G. Amendola 173, 70126 Bari, Italy \\
${}^{(4)}$
Istituto Nazionale di Fisica Nucleare, Sezione di Bari,
Bari, Italy \\
${}^{(5)}$
Relativity and Cosmology Group, 
School of Computer Science and Mathematics,
University of Portsmouth,\\
Portsmouth P01 2EG, England
\par}
%{\small\rm\centering(\ignorespaces February 2001\unskip)\par}

\par
\bgroup
\leftskip=0.10753\textwidth \rightskip\leftskip
\dimen0=-\prevdepth \advance\dimen0 by17.5pt 
\nointerlineskip
\small\vrule width 0pt height\dimen0 \relax

We analyze the signal-to-noise ratio for a relic background of
scalar gravitational radiation composed of massive,
non-relativistic particles, interacting with the monopole mode of two
resonant spherical detectors. We find that the possible signal is 
enhanced with respect to the differential mode of the interferometric
detectors. This enhancement is due to: {\rm (a)} 
the absence of the signal suppression, for non-relativistic scalars, with
respect to a background of massless particles, and {\rm (b)} for flat
enough spectra,  a growth of the signal with the observation time faster
than for a massless stochastic background. 

\par\egroup
\vskip2pc]
\thispagestyle{plain}

\endgroup

The detection of a background of relic gravitational radiation is
undoubtedly one of the main challenges of the physics of the XXI
century. The detection of, or even a direct experimental bound on, a
stochastic gravitational background of primordial origin would
considerably improve our knowledge of  the very early
state of the Universe and our understanding of
physical processes at a Planckian energy
scale (see for instance \cite{1} an references therein). 

In view of the present and forthcoming data, soon available from the
cross-correlation of all existing (resonant and interferometric)
gravitational antennas, the signal-to-noise ratio ($SNR$) due to a
stochastic background has been analyzed in full details  (see \cite{2}
an references therein), in particular for a background of
(ultra)relativistic particles, whose spectral distribution in frequency is
identical to their momentum distribution. 

As far as the tensor sector of gravitational radiation is 
concerned, the relativistic assumption is certainly appropriate and
self-consistent. However, it could not be fully appropriate 
in order to analyze the contribution of a possible scalar sector, which
could correspond to a stochastic background of massive particles (for
instance, dilatons \cite{3}), possibly non-relativistic at the time of
present observations. 

A non-relativistic background of  massive scalar waves requires indeed 
a generalization of the standard $SNR$ expression which takes into
account the relevant mass parameter of the energy spectrum in
momentum space \cite{4}. In that case it has been recently shown that,
in spite of the induced suppression factor (with respect to massless
particles) $(p/E)^4 \ll 1$, it is possible to obtain a resonant response
also from the non-relativistic part of the scalar wave spectrum,
provided the mass lies within the sensitivity band of the detectors
\cite{4,5}. Also, if the energy density of the non-relativistic scalar
background is a significant fraction of the critical energy density, the
signal could be detectable by future planned detector configurations,
such as ``Advanced LIGO" \cite{6}.

The analysis of \cite{4,5} refers however to the $SNR$ produced by a 
massive scalar background in the cross-correlation of two
interferometric detectors. In particular, for a non-relativistic
scalar wave, the suppression factor which appears in the pattern
function is due to the 
interaction of the scalar wave with the differential mode of the
interferometers, which is characterized by a $3\times 3$ traceless 
response  tensor. As already pointed out in \cite{5}, such a suppression
may be absent in the monopole mode of resonant spherical detectors,
whose response tensor is proportional to the  $3\times 3$ identity
matrix~\cite{7}. In addition, in the
resonant response of a sphere to  a non-relativistic scalar background,
and for flat enough spectra, the infrared cut-off of the $SNR$ integral is 
provided {\em not} by the noise  power spectrum of the detector (as
usual in the case massless particles), but by a minimal frequency scale
determined by the mass of the background field and by the total
observation time, $T$. As a consequence, the corresponding $SNR$  may
rise with $T$ much faster than usual (depending on the power of the
scalar spectrum). 

The aim of this paper is to present, and briefly discuss, these two
effects (i.e., the  absence of non-relativistic suppression, and the 
unconventional $T$-dependence of $SNR$) which are new (to the best of
our knowledge), and which could greatly favour the detection of a
non-relativistic scalar background through the correlation of two (or
more) resonant spheres.

We start by recalling the general expression for the optimized $SNR$
obtained through the cross-correlation of two detetctors, with
one-sided noise spectral density $P_1(|f|), P_2(|f|)$, and generated by
a stochastic background of scalar particles with mass $m$ and spectral
energy density $\Om(p)$ in momentum space \cite{4,5}:
\bea
&&\qquad
SNR= \left(\frac{H^2_0}{5\pi^2}\right)\,
\Bigg[2\,T \,\int_{0}^{\infty}{dp\over  p^3\,(p^2+\tilde{m}^2)^{3/2}}
\nonumber\\
&& \qquad
\times
\frac{\Omega^2(p)\,\gamma^2(p)}
{P_1(\sqrt{p^2+\tilde{m}^2})\,P_2(\sqrt{p^2+\tilde{m}^2})}\Bigg]^{1/2}. 
\label{1}
\eea
Here $T$ is the total observation time, $\ti m= m/2\pi$, and $H_0$ is the
present value of the Hubble parameter, which appears in the above
equation because the energy spectrum $\rho(p)$ of the background is
expressed in units of critical energy density \cite{1,2}:
\beq
\Omega(p)= {1\over \rho_c} {d \rho\over d \ln p},~~~~~
\rho_c={3H_0^2\,M_P^2\over 8 \pi}
\label{2}
\eeq
where $M_P$ is the Planck mass. Finally, in eq. (\ref{1}), $\ga(p)$ is the
overlap reduction function \cite{1,2}, which accounts for the relative
orientation and location of the two gravitational antennas, and
determines their efficiency in the detection of a given background. For a
plane  wave of momentum  $\vec p= \hat n p$, where $\hat n$ is a unit
vector specifying the propagation direction, $\ga(p)$ is defined by 
\beq
\gamma(p)=\frac{15}{4\pi}\,\int d^2\hat{n}
e^{2\pi i p\hat{n}\cdot(\vec{x}_{1}-\vec{x}_{2})}
F^{1}(\hat{n})F^{2}(\hat{n}) 
\label{3}
\eeq
where the integration is extended over the full solid
angle and the normalisation constant 
has been chosen so that -- in the massless case -- one
obtains $\gamma(p)=1$ for the differential mode of 
two  coincident and coaligned interferometers. Here 
$\vec{x}_1,\vec{x}_2$ are the positions of the centers of mass 
of the two detectors, 
and $F^1, F^2 $  are the so-called antenna pattern functions \cite{1,2},
determined by the wave polarization tensor, $e_{ab}(\hat n)$ (which
depends on the spin-content of the background), and by the detector
response tensor $D_{ab}$ (which
depends on the geometrical shape of the detector). In general:
\beq
F^i= q_i e_{ab}(\hat n) D^{ab} 
\label{4}
\eeq
where $q_i$ is the effective coupling of the background to the
detector, normalized to $1$ for the geodesic coupling to metric
fluctuations \cite{4,5}.

Note that, for a better comparison with experimental observables, we
are choosing ``unconventional" units $h=1$, so that energy and
frequency simply coincide, $E= (p^2 + \mt^2)^{1/2}=f$. Note also that,
for $m=0$, $p=f$ and one recovers the standard 
form of $SNR$ appropriate to a background of massless particles
\cite{1,2}; the usual result for a background of relic gravitons then follows
by inserting into eqs. (\ref{3}), (\ref{4}) the appropriate spin-two
polarization tensor, with $q=1$. 

For a massive scalar wave, however, one should in general distinguish
two different  pattern functions, one related to the geodesic coupling
of the detector to the scalar component of metric oscillations \cite{8},
the other related to the direct, non-geodesic coupling of the scalar
charge $q_i$ of the detector to the spatial gradients of the scalar
background \cite{9}. By using the transverse and longitudinal
decomposition of the polarization tensor of the massive scalar wave
with respect to the propagation direction $\hat n$, and defining
\beq
T_{ab}=(\delta_{ab}-\hat{n}_a\,\hat{n}_b), ~~~~~
L_{ab}=\hat{n}_a\,\hat{n}_b \,,
\label{5}
\eeq
the two patter functions can be written, respectively, as \cite{5}
\beq
F(\hat n)=D^{ab}\left(T_{ab}+
\frac{\tilde{m}^2}{E^2}\,L_{ab}\right), ~
F_q(\hat n)= q \frac{p^2}{E^2} D^{ab}
\,L_{ab}.
\label{6}
\eeq
The first one applies when considering the response of the detetctor to
the spectrum of scalar metric fluctuations induced by the scalar
background, the second one when considering the direct response of
the detector to the fluctuations of the background itself. For the
differential mode of an interferometric antenna, in particular,
$D^{ab}\da_{ab}=0$ and, in both cases, the pattern function of a massive scalar
wave turns out to be proportional to the massless one, but with the
strong suppression factor $(p/E)^2$ for non-relativistic modes \cite{5}. 

A resonant sphere, on the contrary, has a monopole mode
characterized by the response tensor $D^{ab}=\da^{ab}$ \cite{7}, so
that $D^{ab}L_{ab}=1$, $D^{ab}T_{ab}=2$. The non-geodesic part of the
corresponding pattern function, $F_q(\hat n)$, is still suppressed for all
non-relativistic modes.  For the geodesic part, however, the angular
dependence completely disappears,
\beq
F(\hat n)={3 \mt^2 +2 p^2\over \mt^2 +p^2},
\label{7}
\eeq
and the overlap function (\ref{3}), for the monopole modes of two
spheres, takes the simple form 
\beq
\ga(p)={15\over 2\pi}\left(3 \mt^2 +2 p^2\over \mt^2 +p^2\right)^2
{\sin (2\pi p d)\over pd},
\label{8}
\eeq
where $d$ is the spatial distance between the detectors. The response
to non-relativistic modes ($p\ll m$) is no longer suppressed (actually,
their response is enhanced by the factor $9/4$). 
This is a first interesting result, which already at this level shows
that a resonant sphere is  a promising device for the detection of a
non-relativistic background of massive scalar particles. 

The second interesting result follows from the study of the integral
(\ref{1}) for the $SNR$ associated to the monopole modes of two spheres,
and induced by a cosmic background whose spectrum $\Om(p)$ is
dominated by the non-relativistic sector $p\laq m$. Such a relic scalar
spectrum is naturally expected in string cosmology models of the early
Universe \cite{10} and, as discussed also in \cite{5}, it can be simply
approximated by the power-law behaviour
\beq
\Om(p)=\da ~\Om_0\left(p/ \mt\right)^\da, ~~~
p \laq \mt, ~~~ \da>0,
\label{9}
\eeq
where $\Om_0$ is the present fraction of critical energy density stored
in the scalar background. The low energy branch of the spectrum is
assumed to be non-decreasing ($\da >0$), to guarantee a finite energy
density, \beq
\int_0^{\mt} d \ln p ~\Om(p)=\Om_0.
\label{10}
\eeq
More complicated distributions are also possible, with $\Om(p)$ peaked
at a scale $p_m<\mt$ (see \cite{5,10}), depending on the details of the
cosmological mechanism of production. But the simple
monotonic spectrum (\ref{9}) is already appropriate to the illustrative
purpose of this paper, and will be adopted in what follows for all our
estimates of the sensitivity.

Inserting into eq. (\ref{1}) the expression for  
the overlap reduction function given by eq.~(\ref{8}) and the spectrum
(\ref{9}), and assuming (for a resonant response) that $\mt$ lies within
the detector sensitivity band, so that $P_i(\mt)$ are finite, the
resulting $SNR$ would seem infrared divergent for $\da <1$. Indeed, for
$p\ra 0$, $\ga(p) \ra$ const, $P_i \ra$ const, and
\beq
(SNR)^2 \sim \int_0^{\mt} d p ~p^{2\da-3} \sim \left[p^{2\da
-2}\right]_0^{\mt}.
\label{11}
\eeq
Unlike for massless particles, the infrared part of the integral is not
killed by the instrumental noises, whose power spectra $P_i(|f|)$
(appearing in the denominator of eq. (\ref{1})) blow up to infinity when
their argument approaches zero. In our case, when $p \ra 0$,  $P_i$ 
keep frozen at the frequency scale fixed by the mass of the
background.

It must be noticed, however, that the presence of a lower bound fixed
by $m$ in the allowed range of frequencies, $f =(p^2+\mt^2)^{1/2}\geq
\mt$, toghether with the existence of a ``maximal" finite time scale,
determined by the observation time $T$, necesssarily implies an
intrinsic infrared cut-off for the momentum variable in the $SNR$
integral. Indeed, for a given $T$, the minum observable frequency
interval
\beq
\Da f =(p^2+\mt^2)^{1/2}-\mt > T^{-1},
\label{12}
\eeq
defines, in the limit $p \ra 0$, the minimum momentum scale
\beq
p_{\rm min}= \left(2 \mt /T\right)^{1/2},
\label{13}
\eeq
to be used as the effective lower bound in the $SNR$ integral. This
instrinsic cut-off may thus introduce an important $T$-dependence in
the final value of $SNR$. 

In order to illustrate this effect, and to provide some 
estimates, let us assume for simplicity that the instrumental noises
$P_1$ and $P_2$ are equal, and nearly constant, for $p$ ranging from
$0$ to $\mt$ (i.e., for frequencies between $\mt$ and $\sqrt 2 \mt$):
$P_1(\mt)=P_2(\mt)\equiv S_h(\mt)$. Also, we shall assume that the
distance between the two spheres is negligible. The $SNR$, for
a background of scalar metric fluctuations charaterized by the spectral
distribution (\ref{9}), interacting with the monopole modes of two
spheres, then reads 
\bea
&&
SNR={3 \da H_0^2 \Om_0\over \pi^2 S_h(\mt)}
\sqrt{2T}\nonumber\\ 
&& 
\times
\left[ \int_{p_{\rm min}}^{\mt}{dp \over p^3(p^2+ \mt^2)^{3/2}}
\left(p\over \mt\right)^{2\da}
\left(3 \mt^2 +2 p^2\over \mt^2 +p^2\right)^4\right]^{1/2}
\label{14}
\eea

For $\da \leq 1$ the integral is dominated by the infrared cut-off
$p_{\rm min}$ (see eq. (\ref{11})). We obtain the following estimates: 

\bea
SNR \simeq && {27\over \pi^2}\left(2\over 1-\da\right)^{{1\over 2}}
\left(H_0\over \mt\right)^2{\da ~\Om_0\over \mt~ S_h(\mt)}
\left(\mt T \over 2\right)^{1-{\da\over 2}}, \nonumber\\
&&
\da<1, 
\label{15}
\eea
\bea
SNR \simeq && {27\over \pi^2}
\left(H_0\over \mt\right)^2{ \Om_0\over \mt~ S_h(\mt)}
\left(\mt T \over 2\right)^{{1\over 2}}
\ln \left(\mt T \over 2\right) , \nonumber\\
&&
\da=1. 
\label{16}
\eea
For $\da>1$ the integral is instead dominated by the end point $p=\mt$
of the spectrum, and the $SNR$ reads 
\bea
SNR \simeq && {10\over \pi^2}
\left(H_0\over \mt\right)^2{ \Om_0\over \mt ~S_h(\mt)}
\left(\mt T \over 2\right)^{{1\over 2}}
F(\da), \nonumber\\
&&
\da>1,
\label{17}
\eea
where
\beq
F(\da)=\da \Bigg[\int_0^1 {dx (2x^2+3)^4 x^{2\da-3}\over
(1+x^2)^{11/2}}\Bigg]^{1/2}
\label{18}
\eeq
is a number of order one. The three different cases give results which
are very similar, {\em except for the dependence on the observation
time} $T$, and it is now evident that, for a given $T$, the sensitivity is
strongly enhanced for flat enough spectra ($\da \leq 1$). 

It is important to stress that such a dependence of the $SNR$ on the
total observation time yields an effective enhancement - with
respect to the usual square root law - of the signal, 
for a massive stochastic background, only  
if $m$ is not much smaller than the typical resonant  frequency
$f_0$ of the detector. Indeed, if $m \ll f_0$, the  infrared cut-off
(\ref{12}) is ineffective, being suppressed by a very high instrumental
noise. It is also worth stressing that this enhancement is present
only in the correlation of two spheres. Indeed, only because of their 
particular overlap function (\ref{8}) the $SNR$ integral may
be dominated by the infrared cut-off, and not by the peak of the 
spectrum (as in the case of interferometers). 

For a numerical estimate of the sensitivity we shall now compute
the fraction of critical energy density 
$\Om_0$ required to have a detectable signal, say $SNR >5$. We take 
$T\sim 10^7$ sec as a typical obervation time, and $S_h \sim 10^{-46}
{\rm Hz}^{-1}$ as a possible realistic noise density for a resonant
sphere \cite{11,12} at a frequency $f=
3 \times 10^3$ Hz. The minimal required energy density, for $\da=1/2$,
$\da =1$ and $\da>1$ is given, respectively, by
\bea
&& h_0^2\,\Om_0 \gaq  10^{-8}\left(SNR\over 5.0\right) \left(T\over 10^7 {\rm
sec}\right)^{-{3\over 4}}\left (S_h\over 10^{-46}{\rm Hz}^{-1}\right)
\nonumber \\
&& 
\times \left(\mt \over  3 \cdot 10^3 {\rm Hz}\right)^{9\over
4},~~~~~~~~ \da =1/2,
\label{19}
\eea
\bea 
&& h_0^2\,\Om_0 \gaq  10^{-7}\left[1+0.04 \ln \left(\mt \over  3 \cdot 10^3 {\rm Hz}\right)
\left(T\over 10^7 {\rm
sec}\right)\right]^{-1}
\nonumber \\
&& 
\times 
\left(SNR\over 5.0\right) \left(T\over 10^7 {\rm
sec}\right)^{-{1\over 2}}\left (S_h\over 10^{-46}{\rm Hz}^{-1}\right)
\left(\mt \over  3 \cdot 10^3 {\rm Hz}\right)^{5\over 2},
\nonumber \\
&&
~~~~~~\da =1,
\label{20}
\eea
\bea 
&& h_0^2\,\Om_0 \gaq  {10^{-5}\over F(\da)}\left(SNR\over 5.0\right)
\left(T\over 10^7 {\rm sec}\right)^{-{1\over 2}}
\left (S_h\over 10^{-46}{\rm Hz}^{-1}\right)
\nonumber \\
&& 
\times \left(\mt \over  3 \cdot 10^3 {\rm Hz}\right)^{5\over 2}, ~~~~~~~
\da >1
\label{21}
\eea
where we have used $H_0= 3.2\,h_0\times10^{-18}$ Hz. 

We may note, for comparison, that eq. (\ref{21}), with $\mt$ replaced
by the typical resonant frequency $f_0$, and $F(\da)$ always of order
one (but replaced by a different function of $\da$) also describes the
sensitivity of two spheres to a massless stochastic background, for all
values of the spectral index. Eq. (\ref{21}), with a
different, slightly smaller $F(\da)$, also estimates the sensitivity of
two coincident and coaligned interferometers to the massive spectrum
(\ref{9}), {\em for any} $\da$ (even if the maximal sensitivity is
expected at frequencies lower than $1$ kHz). For the interferometers,
however, the sensitivity is to be further depressed if the spectrum is
peaked at scales smaller than $m$, because of the non-relativistic
suppression factor \cite{5}. For the spherical detectors such a
suppression is absent, and the sensitivity remains at the above
levels independently from the shape of the spectrum. 

We may therefore conclude that the cross-correlation of two resonant
spherical detectors seems to be, in principle, particularly
appropriate to the search for a possible relic cosmic background of
non-reativistic scalar particles, because of two effects:  the
absence of non-relativistic suppression for the monopole mode of the
spheres and, for a flat enough spectrum,  the faster increase of the
sensitivity with the observation time.  The numerical estimates reported
in this paper refer to the case in which the mass of the scalar particle is
within the bandwidth of the antennas.  The bandwidth depends on the
noise performance of the readout system, and is limited to a few Hz in
the currently operating resonant bars \cite{13}. By employing a
xylophone of advanced hollow spheres \cite{12}, or the recently
proposed dual spheres \cite{14}, it should be possible however to open
up the bandwidth and scan a wide mass range at a high level of
sensitivity, comparable with the present numerical estimate, thus
providing new important information on possible ultralight scalar
partners of the gravitons,  and on fundamental string/M-theory models
of the early Universe.

% \acknowledgments
%It is a pleasure to thank ...


\begin{references}
\newcommand{\bb}{\bibitem}

\bb{1}M. Maggiore, Phys. Rep. {\bf 331}, 283 (2000). 


\bb{2}B. Allen and J. D. Romano, Phys. Rev. D {\bf 59}, 102001 (1999). 

\bb{3}M. Gasperini,  Phys. Lett. B {\bf 327}, 314 (1994); 
M. Gasperini and G. Veneziano,  Phys. Rev. D  {\bf 50}, 2519 (1994); 
M. Gasperini, in {\em Proc. of
the 12th It. Conference on ``General Relativity and Gravitational
Physics"} (Rome, September 1996), edited by M. Bassan et al. (World
Scientific, Singapore, 1997), p. 181. 

\bb{4}M. Gasperini, Phys. Lett. B {\bf 477}, 242 (2000).

\bb{5}M. Gasperini and C. Ungarelli, {\sl ``Detecting a relic
background of scalar waves with LIGO"}, Phys. Rev. D (2001) (in press), 
gr-qc/0103035.

\bb{6}A. Abramovici et al., Science {\bf 256}, 325 (1992). 

\bb{7}M. Bianchi, M. Brunetti, E. Coccia, F. Fucito and J. A. Lobo, Phys.
Rev. D {\bf 57}, 4525 (1998). 

\bb{8}M. Maggiore and A. Nicolis, 
Phys. Rev. D {\bf 62}, 024004 (2000). 

\bb{9}M. Gasperini, Phys. Lett. B {\bf 470}, 67 (1999).

\bb{10}M. Gasperini, {\em in Proc. of the IX Marcel Grossman Meeting} 
(Rome, July 2000), eds. R. Ruffini et al. (World Scientific, Singapore),
gr-qc/0009098. 

\bb{11}W. W. Johnson and S. M. Merkowitz, Phys.
Rev. Lett. {\bf 70}, 2367 (1993). 

\bb{12}E. Coccia, V. Fafone, G. Frossati, J. Lobo and J. Ortega,
Phys. Rev. D {\bf 57}, 2051 (1998). 

\bb{13}Z. A. Allen et al., Phys.
Rev. Lett. {\bf 85}, 5046 (2000).

\bb{14}M. Cerdonio et al., Phys.
Rev. Lett. {\bf 87}, 031101 (2001).

\end{references}
\end{document}